\documentclass[preprint,12pt]{elsarticle}
\usepackage{amssymb}
\usepackage{amsmath}
\usepackage{subfigure}
\usepackage{placeins}
\usepackage{lineno}

\journal{Journal of Subatomic Particles and Cosmology}

\begin{document}

\begin{frontmatter}

\title{Measurement of the differential t-channel production cross-section of single~top~quarks and top~antiquarks at $\sqrt{s} = $13 TeV with the ATLAS detector}

\author[BUW]{Maren Stratmann} 
\ead{maren.stratmann@cern.ch}
\author{on behalf of the ATLAS collaboration} 

\affiliation[BUW]{organization={University of Wuppertal},
            country={Germany}}

\begin{abstract}
Differential production cross-sections of single top quarks and top antiquarks are measured in proton--proton collisions at a centre-of-mass energy $\sqrt{s} = 13\,$TeV.
The full Run-2 dataset collected by the ATLAS detector at the LHC in the years 2015--2018 is used.
The differential cross-sections are measured as a function of the transverse momentum and absolute rapidity of the top (anti)quark.
The measurement results are compared to predictions obtained from fixed order calculations, different matrix-element event generators and different parton distribution function sets.
The results agree with the theoretical predictions within the measurement uncertainties.
An effective field theory interpretation of the measurement sets constraints on the contribution of the four-fermion operator $O_{Qq}^{3,1}$.
\end{abstract}


\end{frontmatter}
% \linenumbers

\section{Introduction}
\label{sec:intro}
The main production channel for single top quarks at the LHC is the $t$-channel exchange of a virtual $W$ boson.
Measurements of differential production cross-sections can be used to compare the predictions of different matrix-element generators and parton shower programs.
They are also of interest as an input for proton parton density function (PDF) fits and can be used in searches for beyond Standard Model physics in an effective field theory (EFT) approach.
Differential cross-sections for $tq$ and $\bar{t}q$ production are measured at parton level as a function of the transverse momentum $p_\text{T}$ and absolute rapidity $|y|$ of the top quark and top antiquark.
The measurement uses proton-proton collision data collected by the ATLAS detector~\cite{PERF-2007-01} in the years 2015--2018 at a centre-of-mass energy $\sqrt{s} = 13\,$TeV.
These proceedings are based on the analysis presented in Reference~\cite{ATLAS-CONF-2025-011}.

\section{Analysis strategy} \label{sec:strategy}
Potential signal events are selected according to the expected event signature of the $t$-channel production of single top quarks.
Two separate signal regions (SRs) are defined based on the lepton charge, referred to as $\ell^+$ SR and $\ell^-$ SR respectively.
Candidate events must contain exactly one charged electron or muon, missing transverse energy and exactly two jets out of which exactly one must be identified as originating from a $b$-quark.
Further event selection requirements are imposed to reduce the contribution from background processes, since a high purity in signal events is required for the differential analysis. 
The most important tool for the background reduction is a feed-forward neural network (NN) that was trained to distinguish signal and background events.
The NN assigns an output score $D_{nn} \in [0,1]$ to each event, where high (low) $D_{nn}$ scores are assigned to signal (background) events.
A requirement of $D_{nn} > 0.93$ is imposed for candidate events. 
Signal over background ratios of 6.1 and 3.8 are obtained for the $\ell^+$ SR and $\ell^-$ SR respectively.

\begin{figure}[htbp]
\centering
\includegraphics[width=0.65\textwidth]{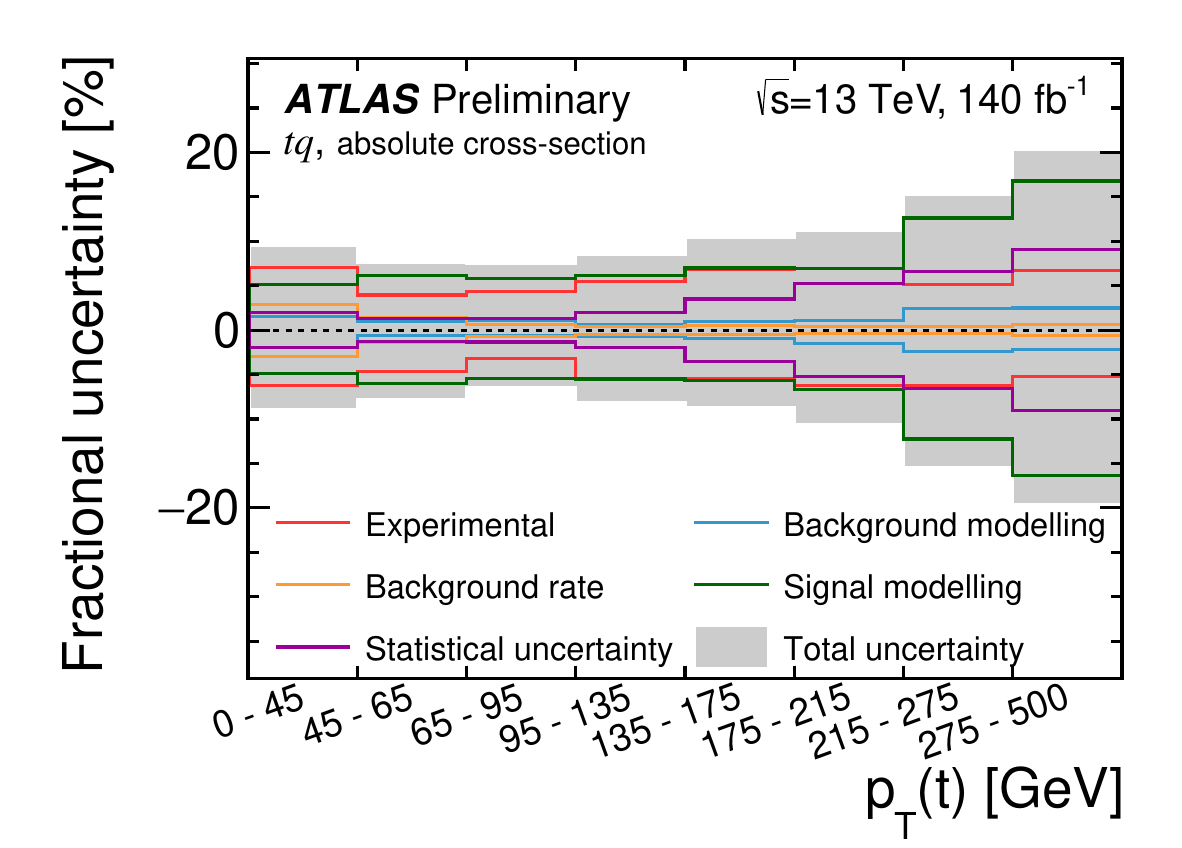}
\caption{
  Systematic uncertainties on the absolute $tq$ production cross-section as a function of $p_\text{T}(t)$~\cite{ATLAS-CONF-2025-011}.
  The shaded band indicates the total uncertainty, obtained by adding all systematic uncertainties in quadrature.
  A further breakdown of the systematic uncertainties into different categories is given by the coloured bands.}
\label{fig:Systs}
\end{figure}

The measured distributions are unfolded to parton level using iterative Bayesian unfolding~\cite{DAgostini:1994fjx}.
The impacts of systematic uncertainties on the unfolded cross-sections are evaluated by unfolding systematically varied distributions with the nominal unfolding corrections and comparing the results to the nominal unfolded distributions.
For systematic variations related to the modelling of the signal process, the unfolded systematically varied distribution is compared to the systematically varied parton-level distribution instead.
The difference between the distributions is defined as the systematic uncertainty in both cases.
All individual uncertainties are added in quadrature to obtain the total uncertainty.
The uncertainty on the absolute $tq$ production cross-section is exemplarily shown in Figure~\ref{fig:Systs}.
The dominating uncertainties are related to the modelling of the signal process and experimental uncertainties.
Uncertainties related to the background processes give the smallest contributions.

\FloatBarrier
\section{Results} \label{sec:results}
\begin{figure}[htbp]
  \centering
  \subfigure[]{
    \includegraphics[width=0.45\textwidth]{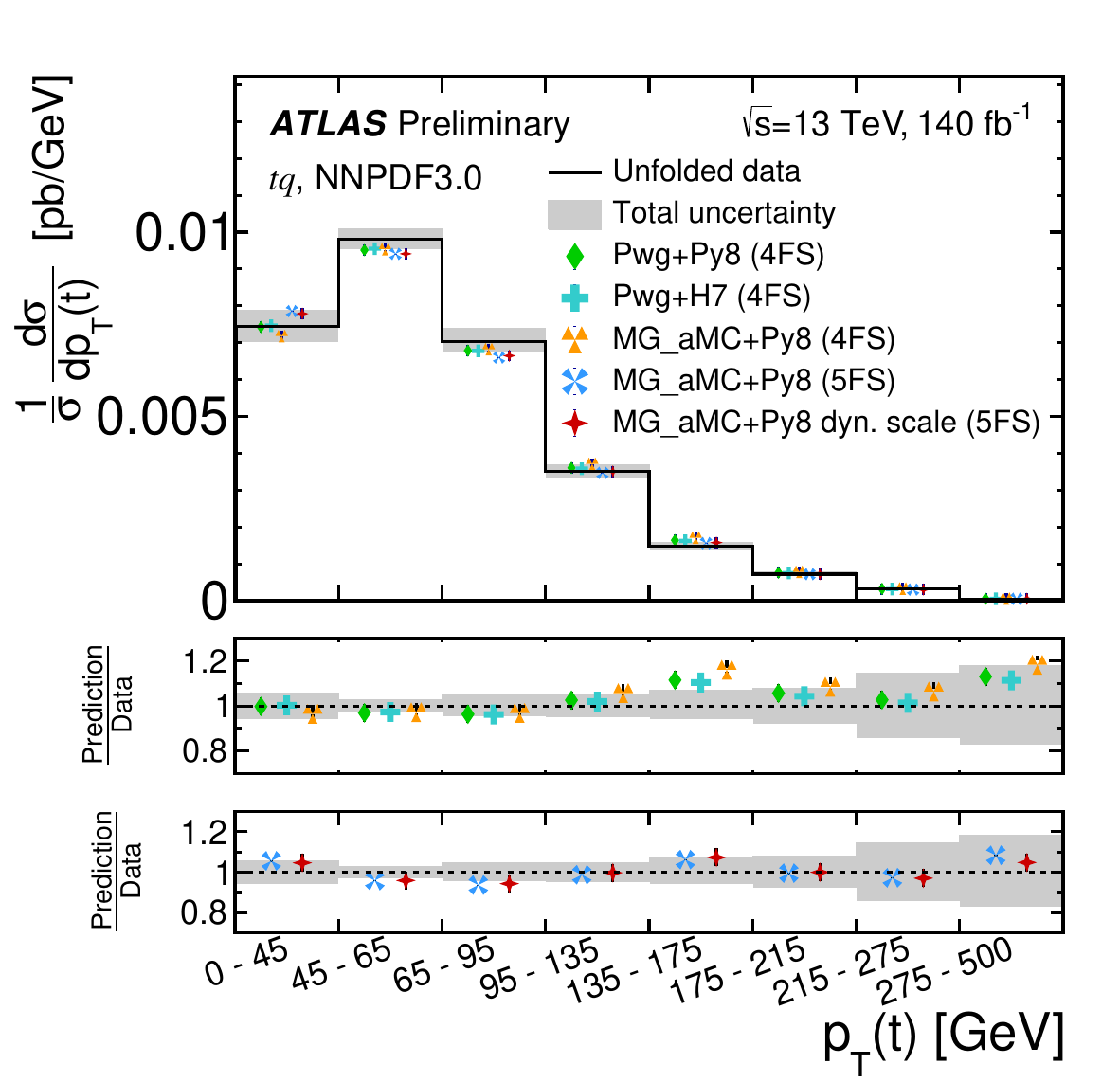}
    \label{fig:Abs}
  }
  \subfigure[]{
    \includegraphics[width=0.45\textwidth]{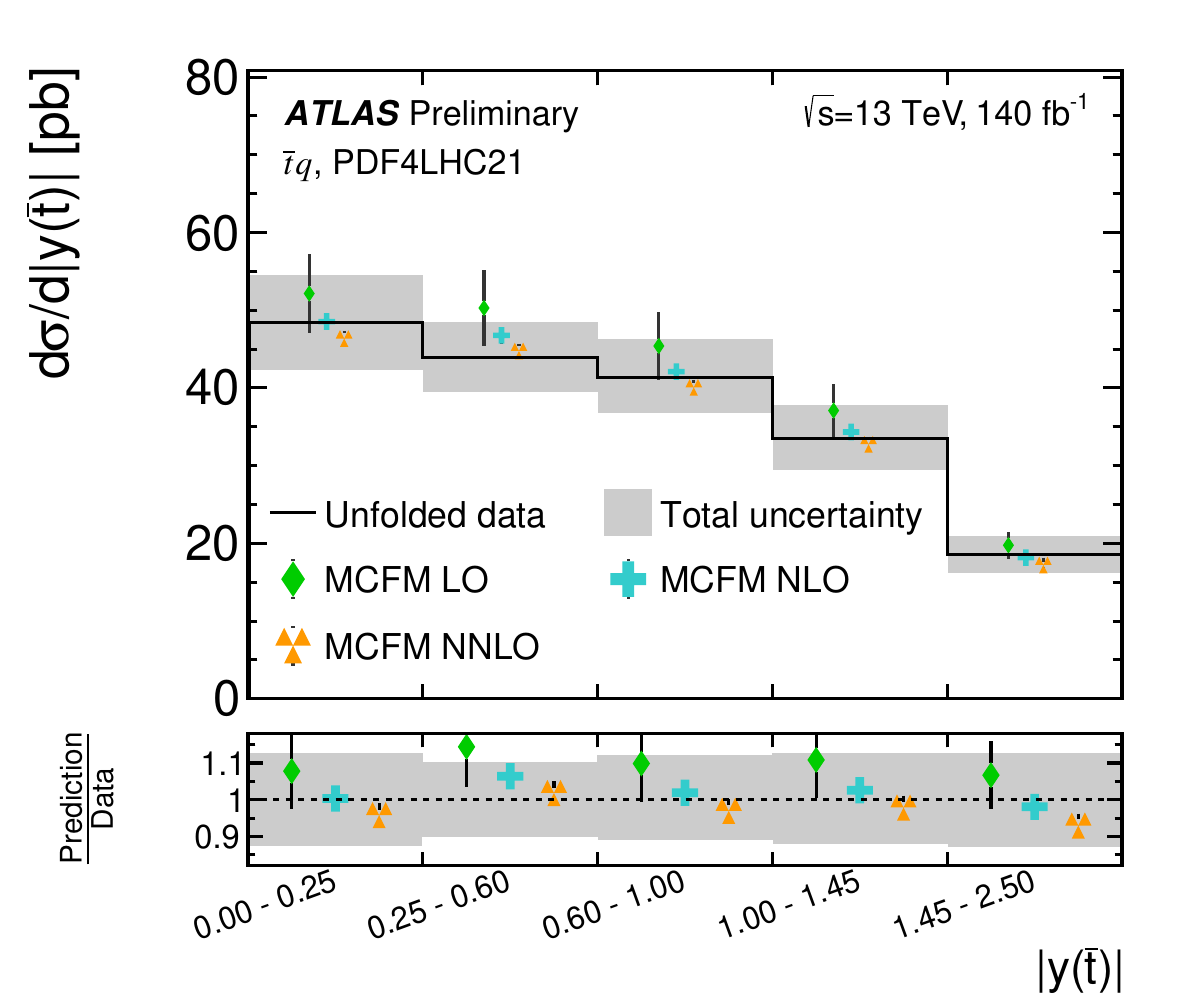}
    \label{fig:Norm}
  } \\
  \subfigure[]{
    \includegraphics[width=0.45\textwidth]{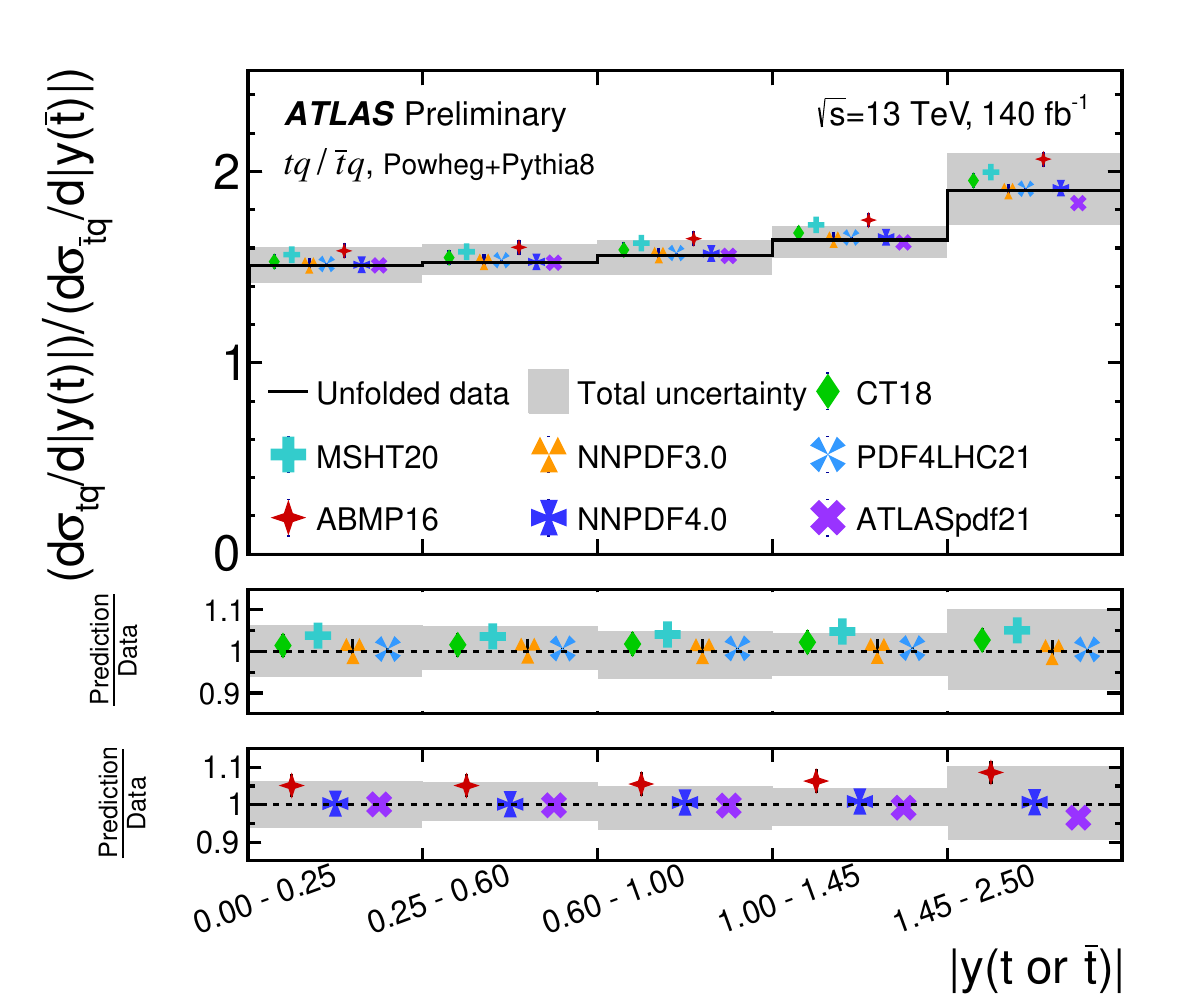}
    \label{fig:Ratio}
  }
  \caption{Measurement results for differential cross-sections compared to theoretical predictions~\cite{ATLAS-CONF-2025-011}.
  \protect\subref{fig:Abs} shows the normalised differential $tq$ cross-section as a function of $p_\text{T}(t)$ compared to theoretical predictions from different matrix-element generators and parton shower programs.
  \protect\subref{fig:Norm} shows the absolute differential $\bar{t}q$ cross-section as a function of $|y(\bar{t})|$ compared to fixed order calculations done with MCFM~\cite{Campbell:2020fhf,Campbell:2021qgd} at different orders in QCD.
  \protect\subref{fig:Ratio} shows the ratio of $tq/\bar{t}q$ cross-sections as a function of $|y(t\text{ or }\bar{t})|$ compared to theoretical predictions calculated with different PDF sets.
  }
  \label{fig:Results}
\end{figure}

The results obtained for the normalised $tq$ cross-section as a function of $p_\text{T}(t)$, the absolute $\bar{t}q$ cross-section as a function of $|y(\bar{t})|$ and 
the ratio of the cross-sections as a function of $|y(t\text{ or }\bar{t})|$ are shown in Figure~\ref{fig:Results}.
The measured cross-sections are compared to various theoretical predictions.
Generally, a good agreement is observed between the predictions and the measured results within the measurement uncertainties.
The ratio of $tq/\bar{t}q$ cross-sections is measured for the first time.

\FloatBarrier
\subsection{Interpretation of the measurement} \label{sec:interpretation}
\begin{figure}[htbp]
\centering
\includegraphics[width=0.65\textwidth]{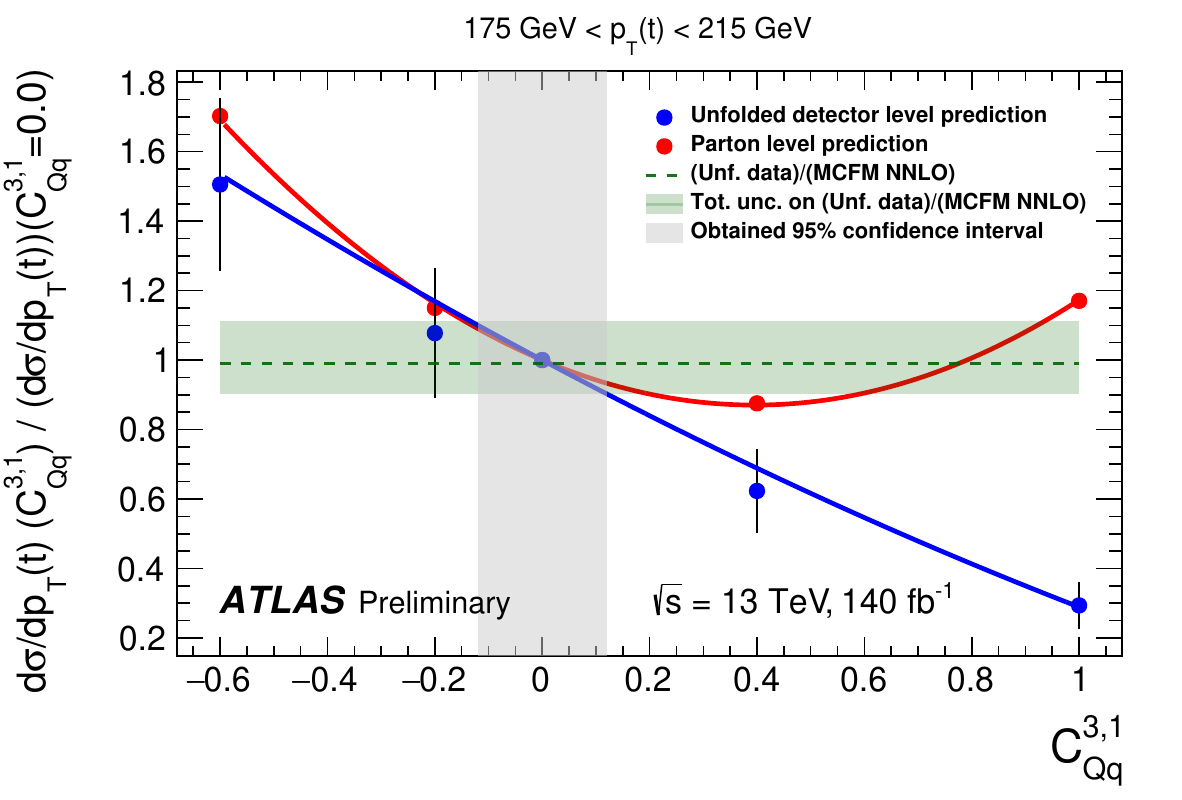}
\caption{
  Parameterisations for the expected relative change of the differential $tq$ production cross-section as a function of $C_{Qq}^{3,1}$ obtained 
  from parton level predictions and by unfolding the detector level predictions~\cite{ATLAS-CONF-2025-011}.
  The uncertainties in the predictions are statistical uncertainties only. 
  The dashed line indicates the measurement result over the theoretical prediction calculated with MCFM at NNLO and the uncertainty band 
  includes the total uncertainty in the measurement result as well as the uncertainty in the theory prediction.
  The vertical shaded area shows the obtained 95\% confidence interval.
  }
\label{fig:EFT}
\end{figure}

The measured differential cross-sections are interpreted in an EFT approach to constrain the contribution of the four-quark operator $O_{Qq}^{3,1}$~\cite{Buckley:2016}.
Simulated samples of the signal process with the associated Wilson coefficient $C_{Qq}^{3,1}$ set to different values are used for the interpretation.
The energy scale $\Lambda$ is set to $1\,$TeV.
The detector level EFT distributions are unfolded to parton level with the nominal unfolding corrections and the unfolded results are used to 
parameterise the expected relative change in the differential cross-section as a function of $C_{Qq}^{3,1}$.
The approach is chosen as non-zero contributions from $C_{Qq}^{3,1}$ alter the selection efficiency of the signal events, which impacts the expected dependency of the differential cross-section on $C_{Qq}^{3,1}$.
The magnitude of the effect is exemplarily shown in Figure~\ref{fig:EFT} for one bin of the differential $tq$ cross-section binned in $p_\text{T}(t)$.
The EFTfitter tool~\cite{EFTfitter} is used to constrain $C_{Qq}^{3,1}$ in a Bayesian statistical framework.
The obtained interval at the 95\% confidence level is
\begin{equation*}
  -0.12\,\text{TeV}^{-2} < C_{Qq}^{3,1}/\Lambda^2 < 0.12\,\text{TeV}^{-2}.
\end{equation*} 

\FloatBarrier

\section{Conclusion} \label{sec:conclusion}
Differential $tq$ and $\bar{t}q$ cross-sections are measured as a function of $p_\text{T}(t)$ or $p_\text{T}(\bar{t})$ and $|y(t)|$ or $|y(\bar{t})|$ in proton-proton collisions collected with the ATLAS detector at the LHC at a centre-of-mass energy $\sqrt{s} = 13\,$TeV.
The measured distributions are unfolded to parton level utilising Iterative Bayesian Unfolding.
The measurement results are compared to various theory predictions obtained from fixed order calculations, different PDF sets and different matrix-element generators and parton shower programs.
Overall, a good agreement is observed within the measurement uncertainties.
The differential ratio $tq/\bar{t}q$ of the cross-sections is measured for the first time.
The results are interpreted in an EFT approach and constraints on the Wilson coefficient $C_{Qq}^{3,1}/\Lambda^2 \in [-0.12,0.12]\,\text{TeV}^{-2}$ are set at the 95\% confidence level.\\

\noindent Copyright 2025 CERN for the benefit of the ATLAS
Collaboration. Reproduction of this article or parts of it is
allowed as specified in the CC-BY-4.0 license

\bibliographystyle{elsarticle-num} 
\bibliography{proceedings.bib}
\end{document}